\documentclass[final,3p,times,twocolumn]{elsarticle}
 \biboptions{comma,sort&compress}
\usepackage{here}
 \usepackage{graphicx}
  \usepackage{epsfig}

\usepackage{amsmath}
\usepackage{amssymb}
\usepackage{slashed}

\def\nin{\noindent}
\def\beq{\begin{equation}}
\def\eeq{\end{equation}}
\def\bea{\begin{eqnarray}}
\def\eea{\end{eqnarray}}

\journal{Nuc. Phys. (Proc. Suppl.)}

\begin{document}

\begin{frontmatter}



\title{Thermodynamic instabilities in nonlocal NJL models.}

 \author[label1,label2]{Marcelo Loewe}\ead{mloewe@fis.puc.cl}
  \address[label1]{Facultad de F\'isica, Pontificia Universidad Cat\'olica de Chile, Casilla 306, Santiago 22, Chile.}
	\address[label2]{Centre for Theoretical Physics and Mathematical Physics, University of Cape Town, Rondebosch 7700, South Africa.}

 \author[label1]{Federico Marquez\corref{cor1}}\ead{cfmarque@uc.cl}
 \cortext[cor1]{Speaker}
  
 \author[label3]{Cristi\'an Villavicencio}\ead{cvillavi@uc.cl}
 \address[label3]{Universidad Diego Portales, 
Casilla 298-V, Santiago, Chile.}
 


\begin{abstract}
\noindent
It has been recently pointed out, that nonlocal Nambu--Jona-Lasinio models, may present unphysical thermodynamical behavior like negative pressure and oscillating entropy. Here we show how these thermodynamic instabilities can be related to the analytical structure of the poles of the quark propagator in the model. The analysis is carried out for two different regulators and we show, in each case, how the instabilities are related to the pressence of highly unstable poles. We also argue that the softening of these instabilities by the inclusion of the Polyakov loop is related to the effect the latter has on the poles of the propagator.
\end{abstract}


\end{frontmatter}


\section{Introduction}
\nin
Nonperturbative QCD and the description of the QCD phase diagram are topics of intense research. Among the different approaches to nonperturbative QCD, the use of effective models has proven to be a powerful tool in studying thermodynamics and the phase diagram. Particularly, the Nambu--Jona-Lasinio (NJL) model \cite{Buballa01, Klevansky01} and its nonlocal version (nNJL) \cite{Birse01, Birse02} have been extensively used for studying thermodynamics of the low energy limit of QCD.\\

In a few recent articles \cite{Blaschke01, Blaschke02}, it has been shown that some odd thermodynamical behavior may occur when working with nNJL models. Negative pressure and oscillating entropy are some of the problems encountered in these cases. This kind of behavior is refered to as {\em thermodynamic instabilities}. In this article, we show that the thermodynamic instabilities are produced by the pressence of some poles in the light quark propagator. The quasiparticle interpretation of the poles of the propagator allows us to comment on the physical meaning of these poles and their relation to the thermodynamic instabilities.\\

Reference \cite{Blaschke02} also shows that the inclusion of the Polyakov loop in nNJL models, produces a softening effect on the thermodynamic instabilities. Here, we consider the effect of the Polyakov loop on the poles of the light quark propagator, in order to understand the reason behind the softening of thermodynamic intabilities.\\

\section{Thermal nonlocal NJL model.}
\nin

We start by considering the nonlocal NJL model described by the Euclidean Lagrangian
\begin{equation}\mathcal{L}_E=\left[\bar{\psi}(x)(-i\slashed{\partial}+m)\psi(x)-\frac{G}{2}j_a(x)j_a(x)\right],\end{equation}
where $\psi(x)$ is a light quark field of bare mass $m$. Nonlocality is incorporated through the currents
\begin{equation}j_a(x)=\int d^4y\,d^4z\;r(y-x)r(z-x)\bar{\psi}(y)\Gamma_a\psi(z),\label{Mcurrent}\end{equation} 
where $\Gamma_a=(1,i\gamma^5\vec{\tau})$. The function $r(x)$ in (\ref{Mcurrent}) is called the {\em regulator} of the model. A bosonization procedure is usually performed through the introduction of a scalar ($\sigma(x)$) and pseudoscalar ($\pi(x)$) field, and the mean field approximation is taken \cite{Scoccola01, Scoccola02, Yo01}. The light quark propagator in Minkowski space is then
\begin{equation}S_0=i\frac{\slashed{q}+\Sigma(-q^2)}{q^2-\Sigma^2(-q^2)},\end{equation}
where $\Sigma(-q^2)=m+\bar{\sigma}r^2(-q^2)$ and $\bar{\sigma}$ is the mean field value of the scalar field. We will encounter singularities for this propagator at $q^2=\Sigma(-q^2)\equiv\mathcal{M}^2$. In what follows, we will adopt the interpretation
\begin{equation}q^2=\mathcal{M}^2=M^2\pm iM\Gamma,\label{Mid}\end{equation}
with $M$ the constituent mass of the quark and $\Gamma$ its decay width. Although different identifications can be made (e.g. $\mathcal{M}=M\pm i\Gamma/2$), we will restrain to the one in Eq. (\ref{Mid}) for simplicity. We can then distinguish three different kinds of poles in the propagator: {\em Real poles:} Poles with $M^2>0$ and $\Gamma=0$ are real poles that correspond to free (deconfined) particle states. {\em Well defined complex poles:} Poles with $M^2>0$ and $0<\Gamma\ll M$ are interepreted as confined quasiparticles with a finite decay width. {\em Ill defined complex poles:} Poles with $M^2<0$ or $\Gamma\geq M$ are objects that cannot be clearly identified with particle states.\\

The usual bosonization procedure and the mean field approximations are then performed \cite{Yo01, Yo02}. We finally arrive to the gap equation
\begin{equation}\frac{\partial\Omega_{MF}}{\partial\bar{\sigma}}=g_0(\bar{\sigma})+\tilde{g}(\bar{\sigma},T)=0,\label{potential}\end{equation}
where
\begin{eqnarray}
g_0(\bar{\sigma})=\frac{\bar{\sigma}}{G}-\frac{N_c}{\pi^2}\int_0^\infty dq_Eq_E^3\frac{r^2(q_E^2)\Sigma(q_E^2)}{q_E^2+\Sigma^2(q_E^2)}\\
\tilde{g}(\bar{\sigma},T)=-\frac{N_c}{\pi^2}\sum_{\mathcal{M}}\left[Z(\mathcal{M}^2)\Sigma(-\mathcal{M}^2)r^2(-\mathcal{M}^2)\right.\nonumber\\
\times\int dkk^2\frac{2n_F(E)}{E}+\left.\left(\mathcal{M}^2\rightarrow(\mathcal{M}^2)^*\right)\right],\label{gfin}
\end{eqnarray}
where $n_F(z)=\left(1+{\rm e}^{\beta z}\right)$ is the Fermi-Dirac distribution, $q_E$ is the momentum in Euclidean space, i.e. $q_E^2=-q^2$ and, as usual, $E=\sqrt{\boldsymbol{q}^2+\mathcal{M}^2}$. In the NJL model $\bar{\sigma}$ and the chiral condensate are related. $\bar{\sigma}(T)$ can be obtained then from Eq. (\ref{potential}) and is an order parameter for the chiral phase transition. This means that the overall thermodynamical information of the model is enclosed on the behavior of $\bar{\sigma}(T)$ or, equivalently, in the poles of the propagator $\mathcal{M}$. Therefore, we will regard any unphysical behavior in $\bar{\sigma}(T)$ as a {\em thermodynamic instability}. Since $\bar{\sigma}(T)$ behaves like the chiral condensate, it should decrease monotonically with temperature so, any rising of $\bar{\sigma}(T)$ with temperature will be considered as a signal of instability. This kind of instability may then produce unphysical behavior like negative pressure and oscillating entropy \cite{Blaschke01, Blaschke02}. $\bar{\sigma}(T)$ will be the main quantity on which we will concentrate in studying thermodynamic instabilities.

\section{Thermodynamic instabilities.}
\nin

Let us start by considering the nNJL model with a Gaussian regulator
\begin{equation}r^2(q_E^2)={\rm e}^{-q_E^2/\Lambda^2},\end{equation}
in Euclidean space. Among the allowed values for the parameters of the model, we consider here two sets chosen to illustrate opposite kinds of analytical structure for the poles of the propagator.

\begin{table}[!h]
\begin{tabular}{|c|c|c|c|c|}
\hline
Set&$\Lambda$(MeV)&$m$(MeV)&$G\Lambda^2$&$\bar{\sigma}_0$(MeV)\\
\hline
A&687&6&28.43&677.8\\
B&1042.2&4.6&15.08&235\\
\hline
\end{tabular}
\caption{Both sets of parameters used for the Gaussian regulator case. $\bar{\sigma}_0$ is the mean value of the scalar field at zero temperature.}
\end{table}

Then, by solving $q^2-\Sigma^2(-q^2)=0$ one can find the poles of the propagator, which means
\begin{eqnarray}
\mbox{Re}(q^2-\Sigma^2(-q^2))&=&0\label{Re}\\
\mbox{Im}(q^2-\Sigma^2(-q^2))&=&0.\label{Im}
\end{eqnarray}

\begin{figure}[!h]
\begin{center}
\includegraphics[width=220pt]{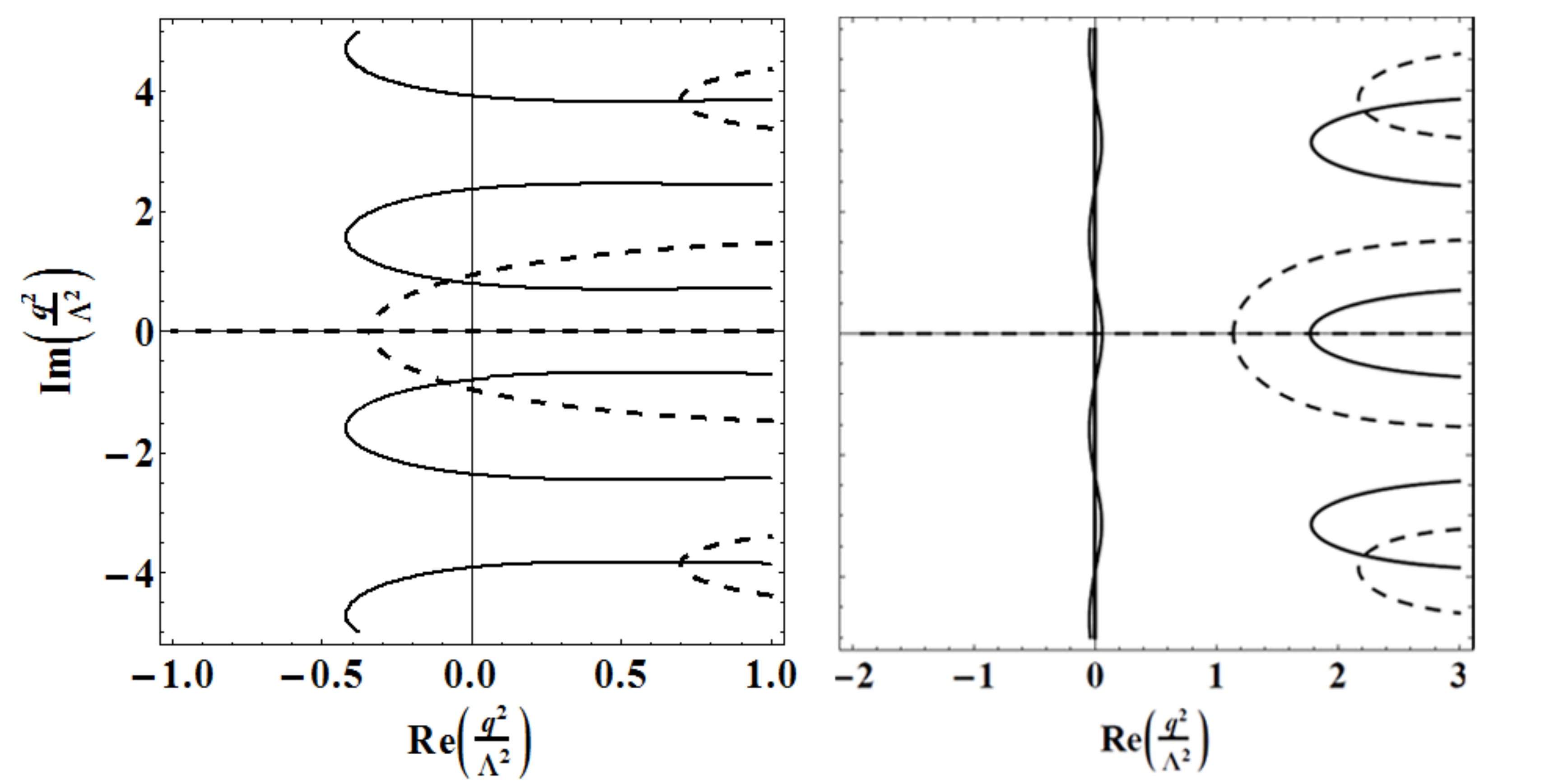}
\end{center}
\caption{Poles of the propagator for set of parameters A (left plot) and B (right plot) for the Gaussian regulator in the $q^2$ plane. The dashed lines are solutions to Eq. (\ref{Im}) and the solid lines solutions to Eq. (\ref{Re}).}
\label{fu}
\end{figure}

Fig. \ref{fu} shows the poles of the propagator for both sets of parameters. Parameter set A exhibits an ill defined complex pole with a negative squared mass followed by an infinite set of ill defined complex poles. Set B, however, exhibits two real poles followed by an infinite set of ill defined complex poles. The infinite number of ill defined complex poles is not very important since, given that they are much more massive than the first poles, their contribution is neglegible. In fact, we will solve the gap equation considering only the first three poles in each case and the result will not differ in any significant way from the complete solution.\\

\begin{figure}[!htb]
\begin{center}
\includegraphics[scale=0.32]{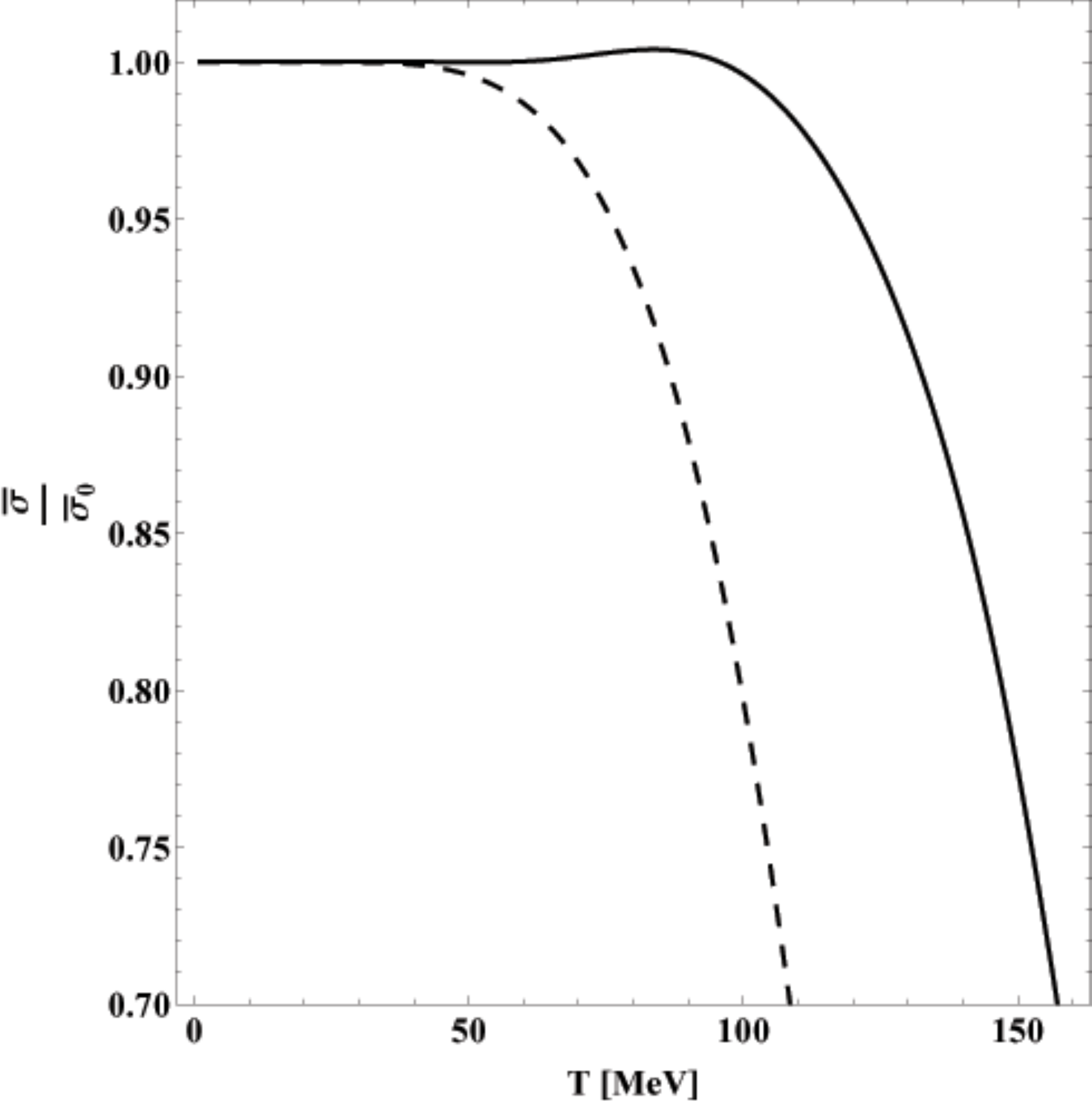}
\end{center}
\caption{Behavior of $\bar{\sigma}$ as a function of temperature. The solid line corresponds to set A and the dashed line to set B. In both cases the the first three poles of the propagator are being considered.}
\label{f3}
\end{figure}

Fig. \ref{f3} shows the behavior of $\bar{\sigma}(T)$ for both sets of parameters. While set B shows no sign of thermodynamic instability, set A presents a bump near $T\approx100$ MeV. As we stated above, all of the thermodynamic information of the model is enclosed in the poles of the propagator. This means that the odd behavior exhibited by $\bar{\sigma}$ must be related to the analytical structure of the poles of the propagator. If one removes the ill defined complex pole of set A, then the instability can be corrected. This will be shown more clearly in our next example.\\ 

Let us now consider a fractional Lorentzian regulator of the form
\begin{equation}r^2(q_E^2)=\frac{1}{1+\left(\frac{q_E^2}{\Lambda^2}\right)^{3/2}},.\end{equation}
in Euclidean space. This regulator is inspired by lattice data from the light quark propagator \cite{Scoccola02}. In order to study the poles of the propagator we need to Wick rotate into Minkowski space. This means we need to define what we will understand by the multivalued function $(-q^2)^{3/2}$, by defining $q^2/\Lambda^2\equiv R{\rm e}^{i\theta}$ and taking $(-q^2/\Lambda^2)^{3/2}=R^{3/2}{\rm e}^{\frac{3}{2}i(\theta+\pi)}$. In this manner we keep the multivalued nature of our functions and the poles of the propagator will occur in two Riemman sheets

\begin{figure}[!htb]
\begin{center}
\includegraphics[width=235pt]{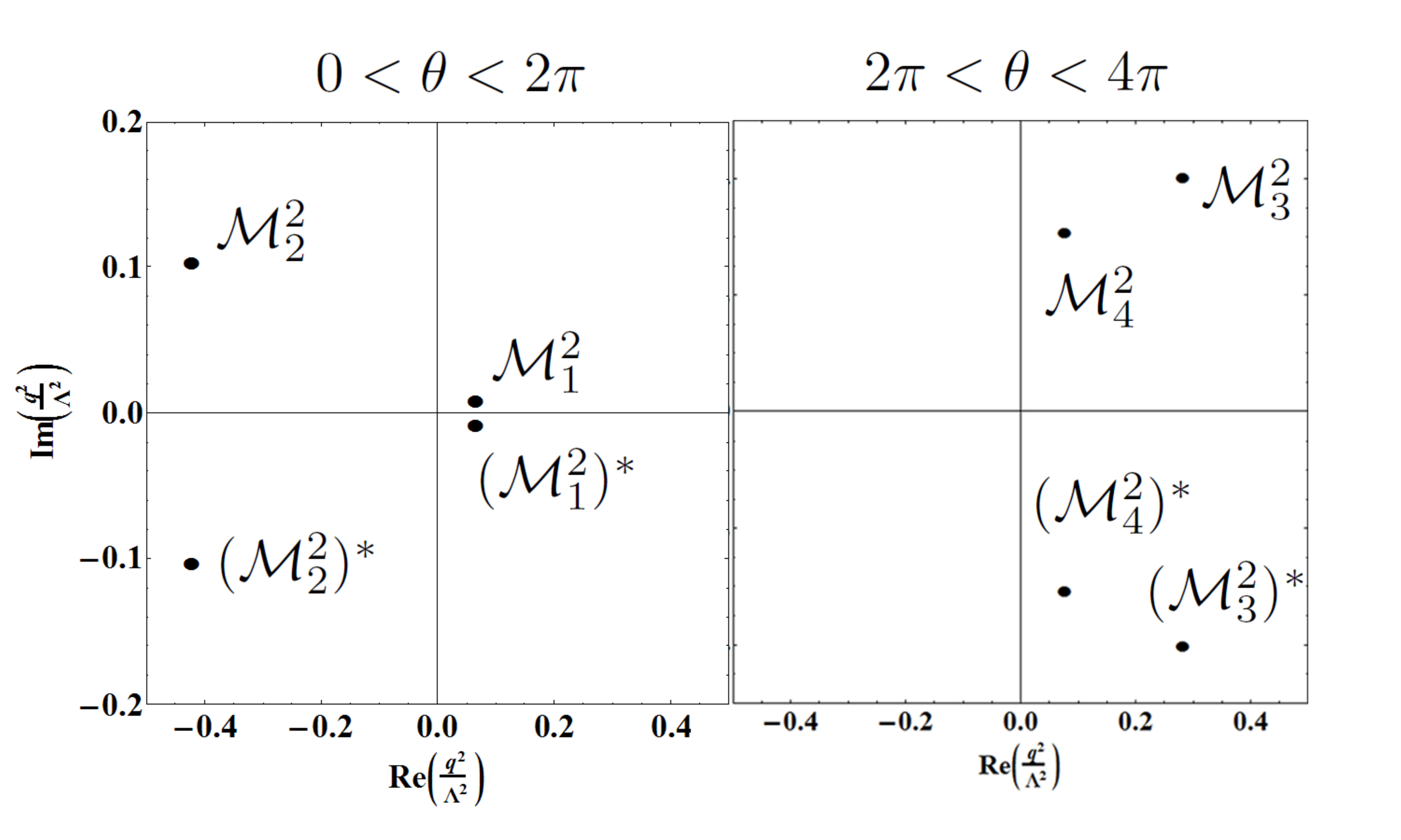}
\end{center}
\caption{Poles of the propagator for the half-integer Lorentzian regulator.}
\label{f8}
\end{figure}

Fig. \ref{f8} shows the poles of the propagator for the fractional Lorentzian regulator. We have four poles: A well defined complex pole $\mathcal{M}_1$ and three ill defined complex poles $\mathcal{M}_{2,3,4}$. One would then expect to have a thermodynamic instability in the model due to $\mathcal{M}_{2,3,4}$. However, if we solve the gap equation considering only $\mathcal{M}_1$ no such instability should appear.

\begin{figure}[!htb]
\begin{center}
\includegraphics[scale=0.31]{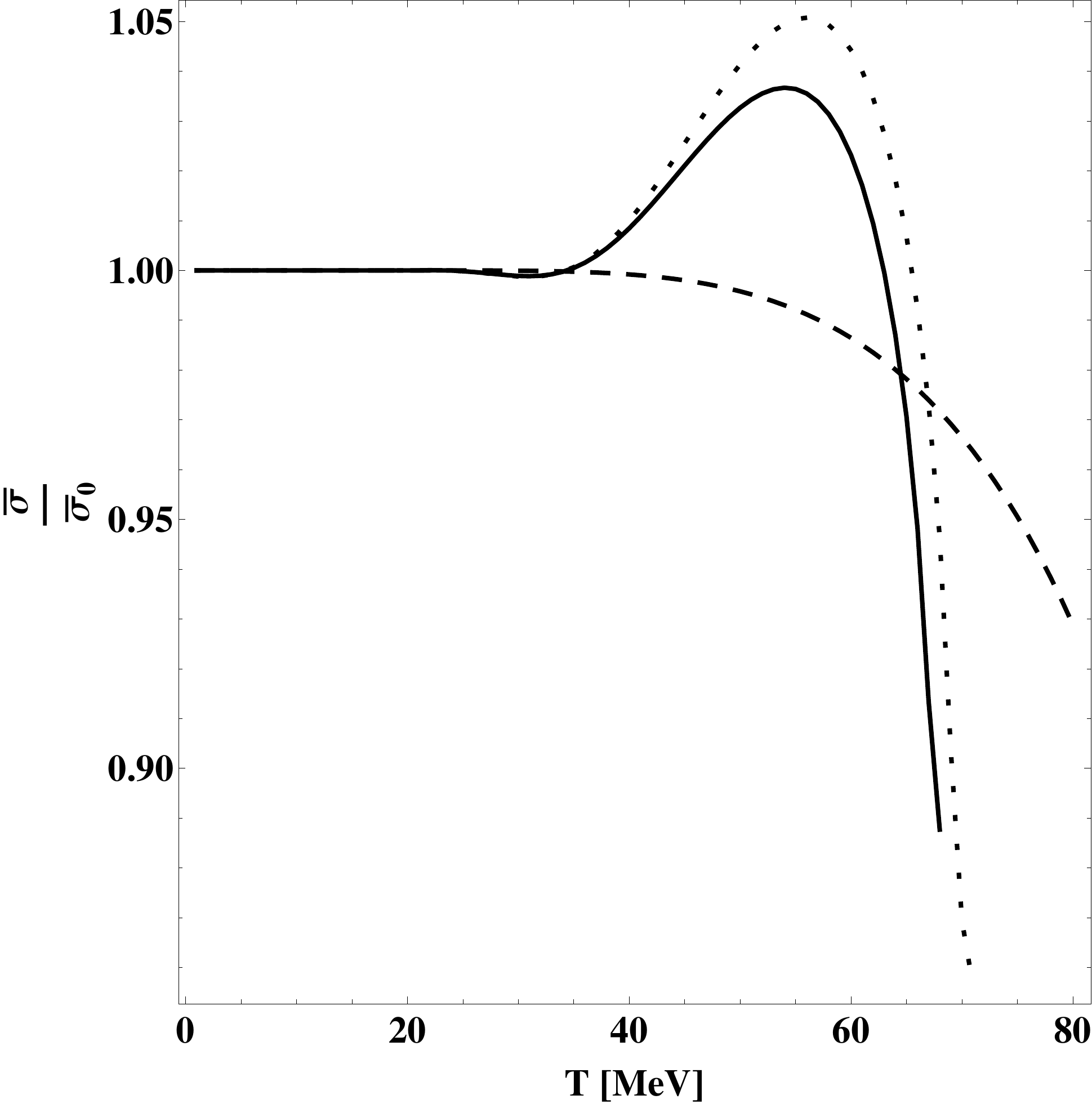}
\end{center}
\caption{Behavior of $\bar{\sigma}$ as a function of temperature. The solid line corresponds to the solution obtained from counting all the poles. The dashed line is the solution from counting only $\mathcal{M}_1$ and the dotted line is the solution from counting only $\mathcal{M}_2$, $\mathcal{M}_3$ and $\mathcal{M}_4$.}
\label{f9}
\end{figure}

Fig. \ref{f9} shows that the instability is present when considering all poles but dissapears once the ill defined complex poles are neglected. Furthermore, the instability grows if only ill defined complex poles are taken into account. We can say then that while ill defined complex poles cause instabilities, real and well defined complex pole exhibit the expected behavior from a condensate.\\

\begin{figure}[!htb]
\begin{center}
\includegraphics[scale=0.2]{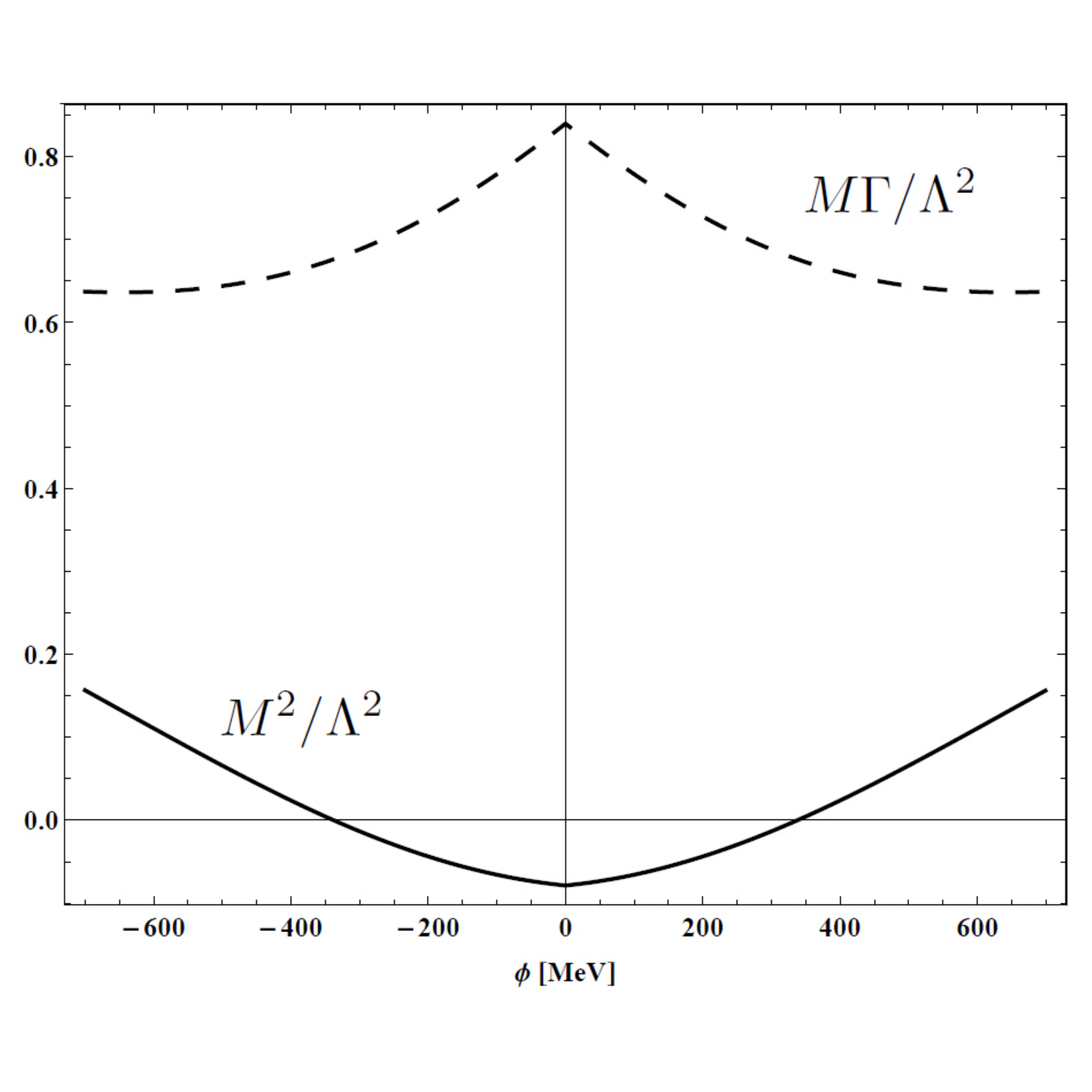}
\end{center}
\caption{Behavior of the of the first pole of the propagator for the Gaussian regulator with parameter set A.}
\label{f11}
\end{figure}

In \cite{Blaschke01} the authors show how these instabilities are softened through the inclusion of the Polyakov loop. Since we have seen how these instabilities are produced, the Polyakov loop should then have a stabilizing effect on the poles in order to soften the instability. In order to see that let us include the Polyakov loop in the model and work in the Polyakov gauge as in \cite{Yo02, Weise01}. The light quark propagator becomes
\begin{multline}S(\phi,q)=\\\frac{\slashed{p}+\Sigma}{(q^2-\Sigma^2(q))\left[(q^2-\Sigma^2(q)-\phi^2/4)^2+q_0^2\phi^2\right]}\boldsymbol{K}\label{propol},\end{multline}
where $\phi$ is the only component of the gluon fields that survives after we specify our gauge and $\boldsymbol{K}$ is a matrix with no singularities whose explicit form is not relevant for us. In Eq. (\ref{propol}) we see that the usual poles of the propagator are still present, but new poles have been included through the second factor in the denominator. Since the Polyakov loop softens these instabilities, the quasiparticles associated with the new poles should be more stable than the previous ones. This is illustrated in Fig. \ref{f11}\\

\section{Conclusions}
\nin

We have discussed the appearance of thermodynamic instabilities in nNJL models with Gaussian and fractional Lorentzian regulators. We showed how these instabilities are related to the presence of ill defined complex poles in the propagator of the model. The thermodynamic instabilities can then be eliminated by simply selecting a set of parameters such that the propagator has no ill defined complex poles (e.g. set B for the Gaussian regulator in this article). This can be considered as a new requirement for the chosen set of parameters. However, it may be the case that such set of parameters cannot be found. In this case one can still eliminate the instabilities by selecting only some poles to be considered in the propagator. This can be done in a well defined manner if one is working in the real time formalism. When calculating the real time thermal propagator from the spectral density function (SDF), the integration contour used to compute the SDF can be modified in order to exclude the ill defined complex poles \cite{Yo02}.\\

We also show how the softening effect of the Polyakov loop, is related to the inclusion of new poles of the propagator together with the Polyakov loop. The quasiparticle states asociated with this new poles are more stable than what is found at vanishing Polyakov loop. Therefore, the thermodynamic instabilities are softened because the contribution from ill defined complex poles to the gap equation is diminished.

\section*{Acknowledgements}
\nin

The authors would like to acknowledge support from FONDECYT under Grant No. 1130056. M.L. also acknowledges support from FONDECYT under Grant No. 1120770. F.M. acknoledges support from CONICYT under Grant No. 21110577.


\end{document}